\documentclass[aps]{revtex4}
\usepackage{amsmath,amssymb,amsfonts}
\usepackage{graphics,graphicx}

\newcommand{\bnabla}{\mbox{\boldmath $\nabla$}}

\begin{document}

\preprint{PRE/Three-body interactions}

\title{Three-body interactions in colloidal systems}

\author{Jure Dobnikar}
 \affiliation{University of Konstanz, Physics Department, 78457 Konstanz, Germany}
 \email{jure.dobnikar@uni-konstanz.de}
\author{Matthias Brunner}
 \affiliation{University of Konstanz, Physics Department, 78457 Konstanz, Germany}
\author{Hans-Hennig von Gr\" unberg}
 \affiliation{University of Konstanz, Physics Department, 78457 Konstanz, Germany}
\author{Clemens Bechinger}
 \affiliation{University of Stuttgart, 2. Physikalisches Institut, 70550 Stuttgart, Germany}

\date{\today}

\begin{abstract}
  We present the first direct measurement of three-body interactions
  in a colloidal system comprised of three charged colloidal
  particles. Two of the particles have been confined by means of a
  scanned laser tweezers to a line-shaped optical trap where they
  diffused due to thermal fluctuations. Upon the approach of a third
  particle, attractive three-body interactions have been observed.
  The results are in qualitative agreement with additionally performed
  nonlinear Poissson-Boltzmann calculations, which also allow us to
  investigate the microionic density distributions in the neighborhood
  of the interacting colloidal particles.
\end{abstract}

\pacs{Valid PACS appear here}
\keywords{Suggested keywords}

\maketitle

\section{Introduction}

Pair interactions in dense systems are in general affected by the
presence of many other surrounding particles. To take such many-body
interactions into account, the degrees of freedom of other particles
are often integrated out, leading to {\sl effective pair potentials}.
This concept is often the only way to handle systems where a large
number of different length and time scales coexist. It is important to
realize, however, that effective potentials - in contrast to {\sl true
  pair potentials} - can not be regarded as fundamental quantities
because their parameters depend on the state of the system. In
addition, no unique way to derive the effective potentials exists and
the effective pair potential picture very often leads to thermodynamic
inconsistencies \cite{Louis}. Accordingly, a correct description of
any liquid or solid must explicitly take into account many-body
effects (and in particular three-body effects as the leading term).
Already in 1943 it has been supposed by Axilrod and Teller (AT)
\cite{Teller} and later also by Barker and Henderson \cite{Barker}
that three-body interactions may significantly contribute to the total
interaction energy in noble gas systems. This seems to be surprising
because noble gas atoms posses a closed-shell electronic structure and
are therefore often (and erroneously) regarded as an example of a
simple liquid . The conjecture of Axilrod and Teller, however, was
confirmed only very recently, when large-scale molecular dynamics
simulations for liquid xenon and krypton \cite{Bomont,Jakse} was
compared with structure factor measurements at small q-vectors
performed with small-angle neutron scattering
\cite{Formisano1,Formisano2}. In these papers it has clearly been
demonstrated that only a combination of pair-potentials and three-body
interactions, the latter in the form of the AT-triple-dipole term
\cite{Teller}, leads to a satisfactory agreement with the experimental
data.  In the meantime, it has been realized that many-body
interactions have to be considered also for nuclear interactions
\cite{Negele}, inter atomic potentials, electron screening in metals
\cite{Hafner}, photo-ionization, island distribution on surfaces
\cite{Osterlund}, and even for the simplest chemical processes in
solids \cite{Ovchinnikov} like breaking or making of a bond.

In view of the general importance of many-body effects it seems
surprising that until now no direct measurements of these interactions
have been performed. This is largely due to the fact that in atomic
systems, positional information is typically provided by structure
factors or pair-correlation functions, i.e. in an integrated form.
Direct measurements of many-body interactions, however, require direct
positional information beyond the level of pair-correlations, which is
not accessible in atomic or nuclear systems. In contrast to that,
owing to the convenient time and length scales involved, the
microscopic information is directly accessible in colloidal
suspensions. In addition, the pair-interactions in colloidal
suspensions can be varied over large ranges, e.g. from short-ranged
steric to long-ranged electrostatic or even dipole-dipole
interactions.

In the present study we used charged colloidal particles whose
interactions are mediated by the microscopic ions in the
electrolyte. The pair interaction in such systems is directly related
to the overlap of the ion clouds (double-layers) which form around the
individual colloids and whose thickness is determined by the ionic
strength of the solution. In highly de-ionized solutions, these
double-layers can extend over considerable distances. If more than two
colloids are close enough to be within the range of such an extended
double layer, many-body interactions are inevitably the
consequence. Accordingly, deviations from pair-wise additive
interaction energies are expected in charge-stabilized colloidal
systems under low salt conditions.  

Here we present the first direct measurement of three-body
interactions, performed in a suspension of charged colloidal
particles. This was achieved by scanned optical tweezers, which
provided a trapping potential for two colloidal particles. When a
third particle was present, considerable deviations from pair-wise
additive particle interactions have been observed. These deviations
increased, as the distance of the third particle was decreased and
were used to extract three-body interaction potentials. We have
additionally performed non-linear Poisson-Boltzmann calculations for
the same parameters and same configurations as chosen in the
experiment. Deriving the interaction potentials from the solutions of
the Poisson-Boltzmann equation, we have correctly taken three-body
terms into account. The numerically obtained three-body potentials are
in qualitative agreement with the experimental results.

Experimental evidence for many-body interactions has been already
obtained from effective pair-interaction potential measurements of
two-dimensional (2D) colloidal systems. Upon variation of the particle
density, a characteristic dependence of the effective pair interaction
was found which has been interpreted in terms of many-body
interactions \cite{gr}. However, during those studies the relative
contributions of different many-body terms could not be further
resolved. Performing the experiment described in this paper, i.e.
observing the system of only three particles, we were able to measure
the three-body interactions directly. 

\section{Experimental system}

As colloidal particles we used charge-stabilized silica spheres with
990nm diameter suspended in water. A highly diluted suspension was
confined in a silica glass cuvette with 200$\mu$m spacing. The cuvette
was connected to a closed circuit, to deionize the suspension and thus
to increase the interaction range between the spheres. This circuit
consisted of the sample cell, an electrical conductivity meter, a
vessel of ion exchange resin, a reservoir basin and a peristaltic pump
\cite{Palberg}. Before each measurement the water was pumped through
the ion exchanger and typical ionic conductivities below 0.07$\mu$S/cm
were obtained. Afterwards a highly diluted colloidal suspension was
injected into the cell, which was then disconnected from the circuit
during the measurements. This procedure yielded stable and
reproducible ionic conditions during the experiments. Due to the ion
diffusion into the sample cell, the screening length $\kappa^{-1}$
decreased linearly with time during the measurements. The rate of
change of the screening length, however, was only less than half a
percent per hour, which means that in the time needed to perform a
complete set of measurements, the ionic concentration did not change
more than about 1 percent. This tiny variation has been taken into
account when performing the PB calculations (see section IV.).\\

\begin{figure}
 \includegraphics[width=0.75\textwidth]{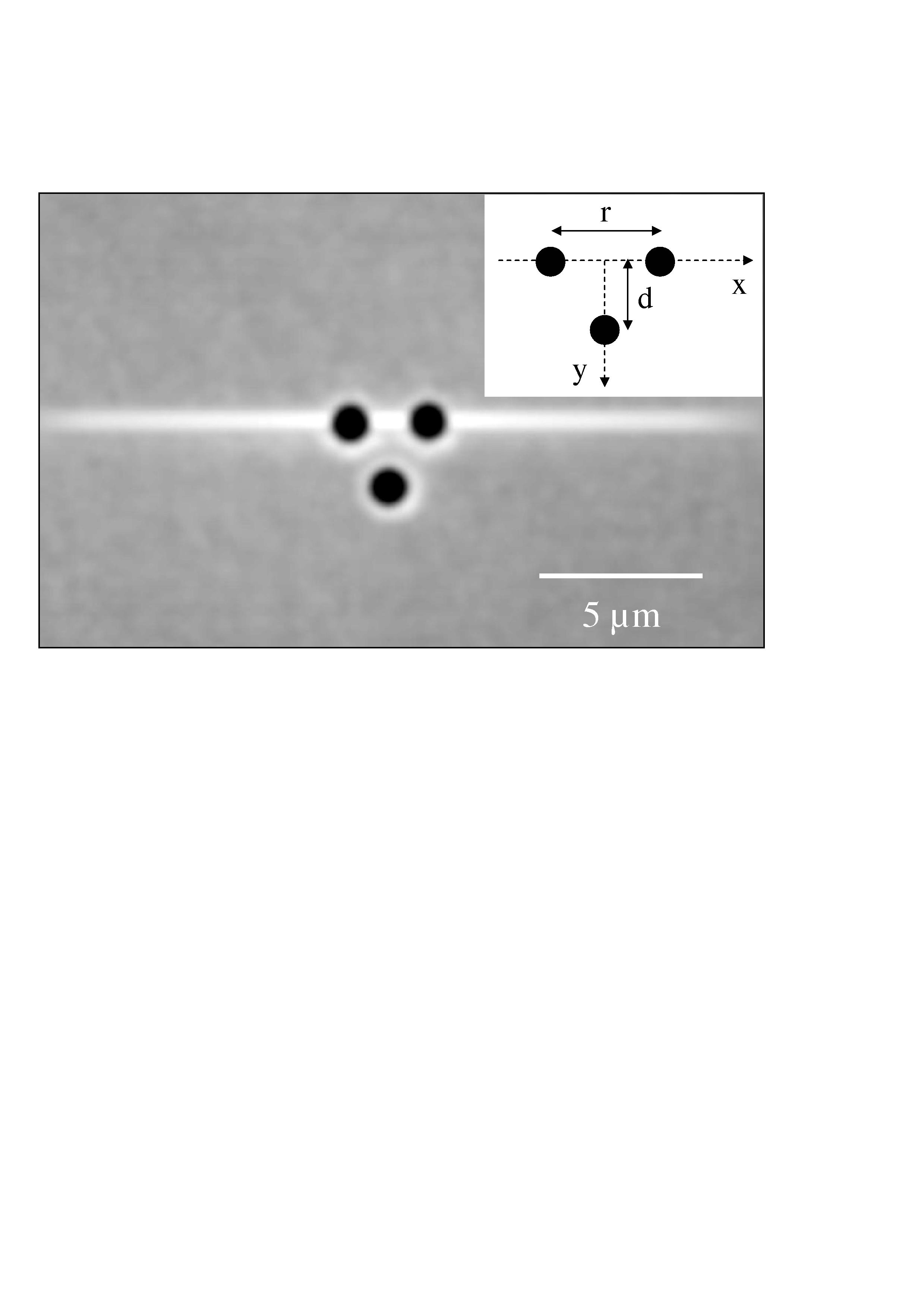}
 \caption{}
\label{photo}
\end{figure}

First, three particles were brought in the field of view of the
microscope after they had sedimented down to the bottom plate of the
sample cell (Fig.\ref{photo}). Two particles were trapped with
line-scanned optical tweezers, which was created by the beam of an
argon ion laser being deflected by a computer-controlled
galvanostatically driven mirror with a frequency of approximately
350Hz.  The time averaged intensity along the scanned line was chosen
to be Gaussian-distributed with the half-width $\sigma_{x}\approx$ 4.5
$\mu$m. The laser intensity distribution perpendicular to the trap was
given by the spot size of the laser focus, which is also Gaussian with
$\sigma_{y}\approx$ 0.5 $\mu$m. This yielded an external laser
potential acting as a stable quasistatic trap for the particles.  Due
to the negatively charged silica substrate, the particles also
experience a repulsive vertical force, which is balanced by the
particle weight and the vertical component of the light force. The
potential in the vertical direction is much steeper than the in-plane
laser potential, therefore vertical particle fluctuations can be
disregarded. The particles were imaged with a long-distance, high
numerical aperture microscope objective (magnification $\times$63)
onto a CCD camera and the images were stored every 120 ms.  The
lateral positions of the particle centers were determined with a
resolution of about 25 nm by a particle recognition algorithm.

Three-body interaction potentials were measured in this setup by
performing the following steps (which will be explained in detail
below): First only one particle was inserted into the trap and its
position probability distribution was evaluated from the recorded
positions. From this the external laser potential $u_{L}$ could be
extracted. Next, we inserted two particles in the trap and measured
their distance distribution. From this, the pair-interaction potential
was obtained. Finally, a third particle was made to approach to the
optical trap by means of additional point optical tweezers (focus size
$\approx$ 1.3 $\mu$m), which held this particle at a fixed position
during the measurement. From the distance distribution of the first
two particles we obtained the total interaction potential for the
three particles. Finally, we substracted a superposition of pair
potentials (known from the previous two-particle measurements) from
the total interaction energy to obtain the three-body interaction.

\section{Data evaluation and experimental results}

\begin{figure}
 \includegraphics[width=0.75\textwidth]{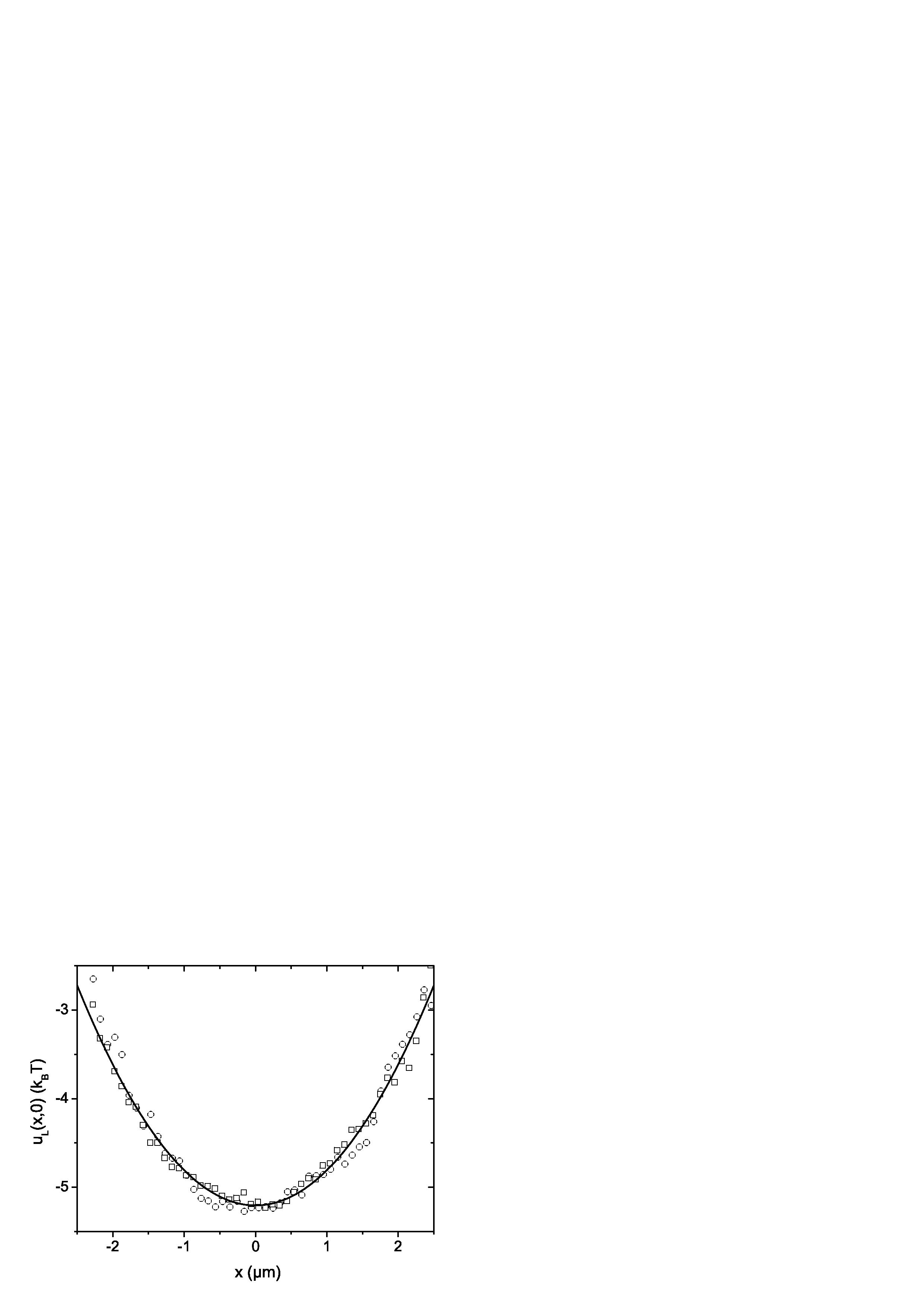}
 \caption{}
\label{laserpot}
\end{figure}

We first determined the external potential acting on a single particle
due to the optical line trap. The probability distribution $P(x,y)$ of
finding a particle at the position $(x,y)$ in the trap was evaluated
from the recorded positions. $P(x,y)$ depends only on the temperature
and the external potential $u_{L}(x,y)$ created by the laser tweezers.
According to the Boltzmann probability distribution
$P(x,y)=P_{L}\rm{e}^{-\beta u_{L}(x,y)}$, with $P_{L}$ being a
normalization constant. Taking the logarithm of $P(x,y)$ yields the
external potential $u_{L}(x,y)$ with an offset given by $\log{P_{L}}$.
The probability distributions in $x$ and $y$ directions are
statistically independent, and can therefore be factorized. The laser
potential is thus $u_{L}(x,y)=u_{L}(x)+u_{L}(y)$. The potential along
the $x$ axis is shown in Fig.\ref{laserpot} for various laser
intensities.  As can be seen, all renormalized potentials fall, within
our experimental resolution, on top of each other. This clearly
demonstrates that the optical forces exerted on the particles scale
linearly with the input laser intensity. This fact allows us to use
different external laser powers for two-body and three-body experiments
(in the three-body experiment, due to the additional repulsion of the
third particle, a stronger laser power is needed to keep the mean
distance between the two particles similar). The corresponding
potential in the perpendicular ($y$) direction has the same (Gaussian)
shape, but it is much steeper due to the chosen scanning direction.
Therefore, the particles hardly move in the $y$ direction during a
measurement.

\begin{figure}
 \includegraphics[width=0.75\textwidth]{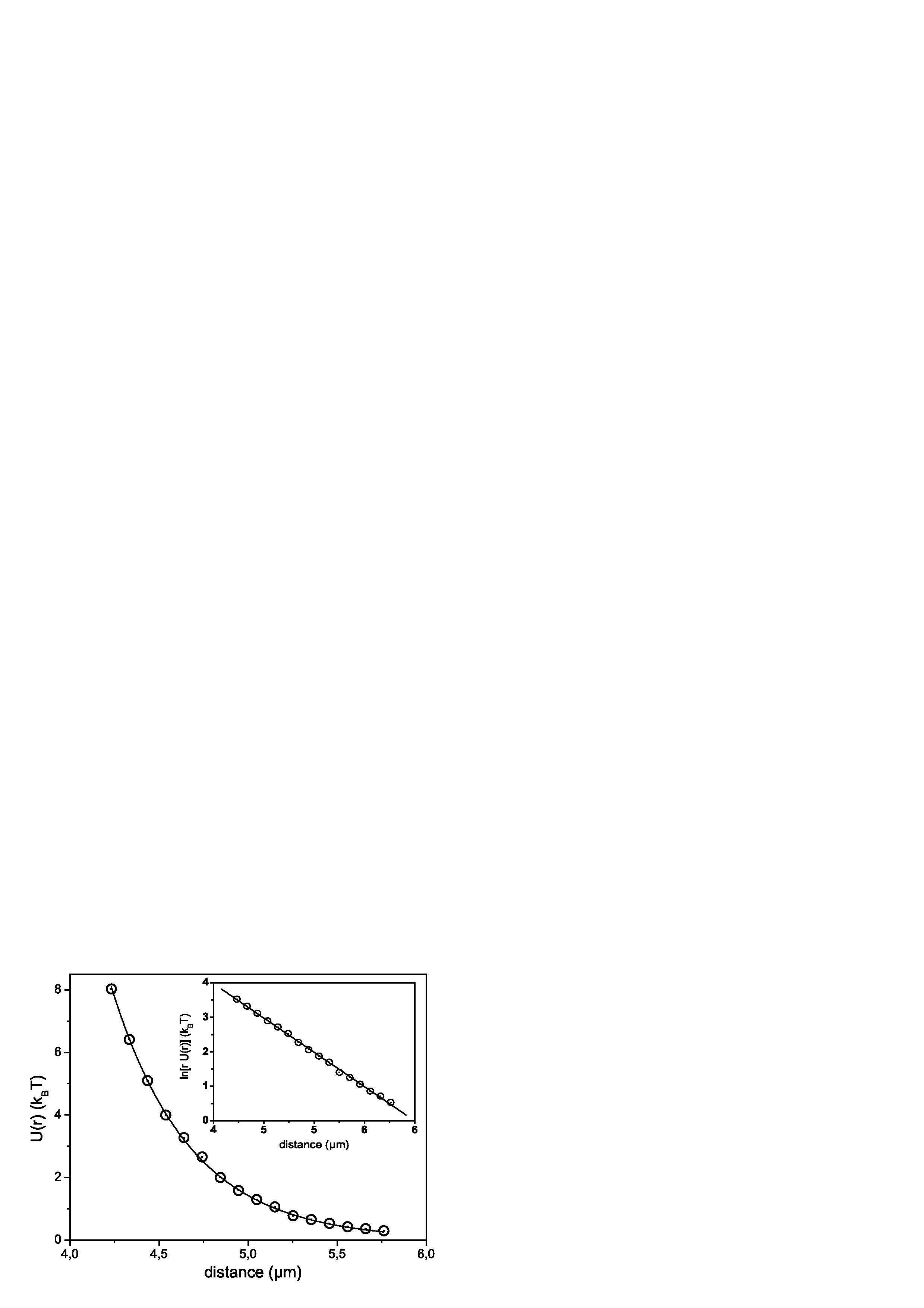}
 \caption{}
\label{pairpot}
\end{figure}

\begin{figure}
 \includegraphics[width=0.75\textwidth]{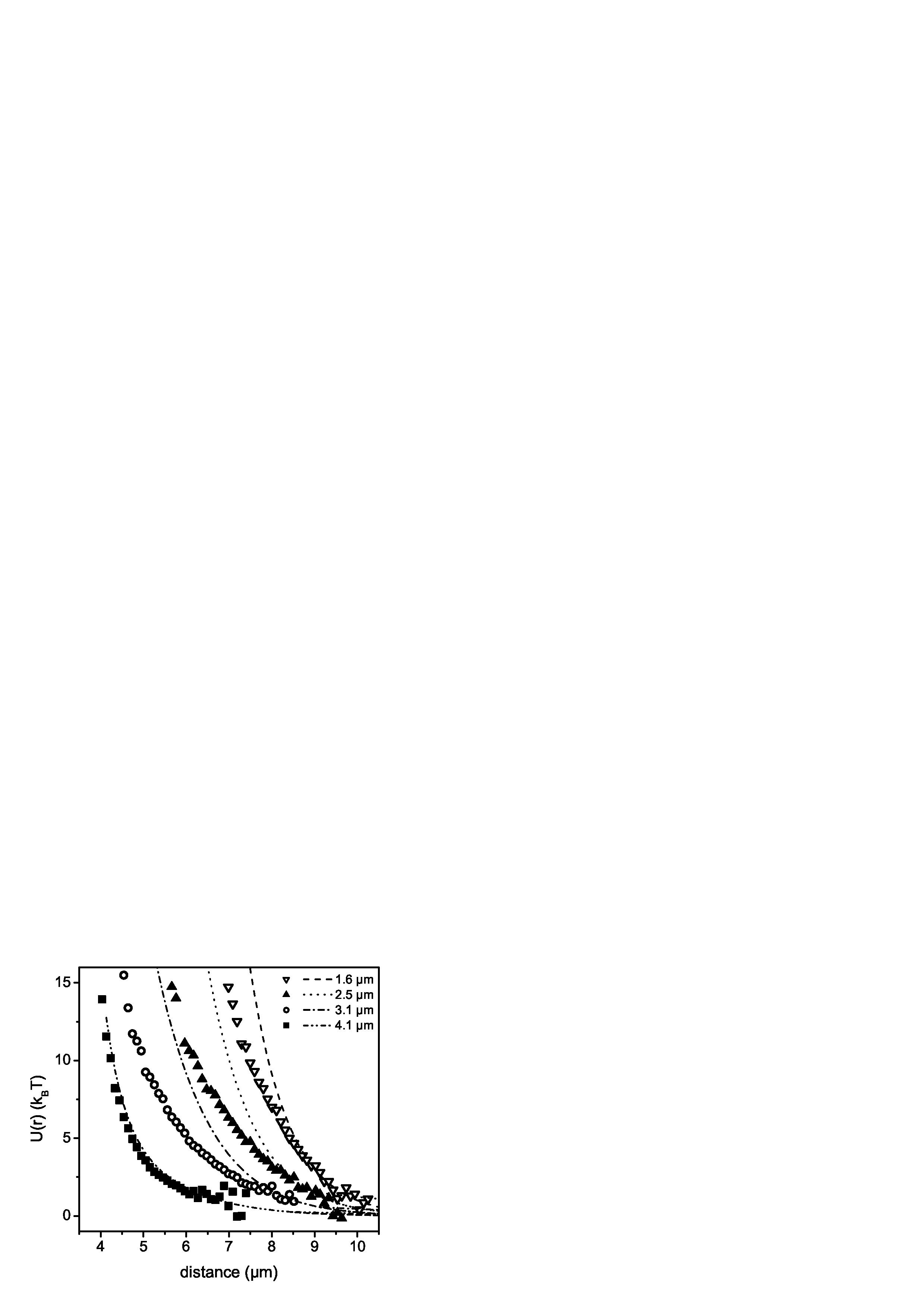}
 \caption{}
\label{utot}
\end{figure}

Next, we inserted a second particle in the trap. The four-dimensional
probability distribution is now
$P(x_{1},y_{1},x_{2},y_{2})=P_{12}\rm{e}^{-\beta \left
(u_{L}(x_{1},y_{1})+u_{L}(x_{2},y_{2})+U(r)\right )}$, with $x_{i}$
and $y_{i}$ being the positions of the $i$-th particle relative to the
laser potential minimum and $U(r)$ the distance dependent
pair-interaction potential between the particles. This can be
projected to
\begin{widetext}
\begin{eqnarray}
P(r)&=&\int\!\!\!\int\!\!\!\int\!\!\!\int P(x_{1},y_{1},x_{2},y_{2})
\delta\!\left (\!\sqrt{(x_{1}-x_{2})^2 + (y_{1}-y_{2})^2}-r\right )
dx_{1}dx_{2}dy_{1}dy_{2} = 
\nonumber \\
&=&P_{12}\rm{e}^{-\beta U(r)}\!\int\!\!\!\int\!\!\!\int\!\!\!\int 
\rm{e}^{-\beta \left (u_{L}(x_{1},y_{1})+u_{L}(x_{2},y_{2})\right )}
 \, \delta\!\left (\!\sqrt{(x_{1}-x_{2})^2 + (y_{1}-y_{2})^2}-r\right ) 
dx_{1}dx_{2}dy_{1}dy_{2}
\label{projection}
\end{eqnarray}
\end{widetext}
In principle the integral is constituted of all possible
configurations of two particles with distance $r$. Performing the full
four-dimensional integration, however, is difficult because of the
limited experimental statistics. This problem can be overcome by the
following two considerations. First, due to the Gaussian shape of the
external potential, the most likely particle configurations are
symmetric with respect to the potential minimum of $u_{L}$ (any
asymmetric configuration for constant $r$ has a higher
energy). Secondly, particle displacements in $y$-direction are
energetically unfavorable because $\sigma_{x} \gg
\sigma_{y}$. Accordingly, for $r = const$ the minimum energy
configuration is $(x_{1}=r/2, y_{1}=0, x_{2}= -r/2, y_{2}=0)$. It has
been confirmed by a simple calculation with the experimental
parameters that all other configurations account for only for less than 1
percent of the value of the integral in
Eq.(\ref{projection}). Accordingly, Eq.(\ref{projection}) reduces to
\begin{eqnarray}
P(r)=P_{0}\rm{e}^{-\beta \left ( U(r)+2u_{L}(r/2,0) \right )} \; .
\label{Pr}
\end{eqnarray}
Since $u_{L}(x,y)$ is known from the previous one-colloid measurement,
we can obtain the interaction potential $U(r)$ from the measured $P(r)$,
\begin{eqnarray}
\beta U(r)\, =\, -\log{P(r)}\, -\, 2\beta u_{L}(r/2,0)\, +\, \log{P_{0}}\;.
\label{Ur}
\end{eqnarray}
The normalization constant $P_{0}$ was chosen in a way that $U(r) \to
0$ for large particle separations $r$.  We first measured $U(r)$
according to the above procedure in the absence of a third
particle. As expected, the negatively charged colloids experience a
strong electrostatic repulsion which increases with decreasing
distance. The pair-interaction potential of two charged spherical
particles in the bulk is well known to be described by a Yukawa
potential \cite{Landau,Overbeek}
\begin{eqnarray}
\beta U(r)=\beta u_{pair}(r)=\left ( Z^{*}\right ) ^2
\lambda_{B} \Biggl (\frac{\rm{e}^{\kappa R}}{1+\kappa R}\Biggr )^2
\frac{\rm{e}^{-\kappa r}}{r} \; ,
\label{Yukawa}
\end{eqnarray}
where $Z^{*}$ is the renormalized charge \cite{Belloni} of the
particles, $\lambda_{B}$ the Bjerrum-length characterizing the solvent
($\lambda_{B}=e^2/4\pi\epsilon\epsilon_{0}k_{B}T$, with $\epsilon$ the
dielectric constant of the solvent and $e$ the elementary charge),
$\kappa^{-1}$ the Debye screening length (given by the salt
concentration in the solution), $R$ the particle radius and $r$ the
centre-centre distance of the particles. Fig.\ref{pairpot} shows the
experimentally determined pair-potential (symbols) together with a fit
to Eq.(\ref{Yukawa}) (solid line). As can be seen, our data are well
described by Eq.(\ref{Yukawa}). As fitting parameters we obtained
$Z^{*}\approx 6500$ electron charges and $\kappa^{-1} \approx 470$nm,
respectively. The renormalized charge is in good agreement with the
predicted value of the saturated effective charge of our particles
\cite{Zsat,Trizac} and the screening length agrees reasonably with the
bulk salt concentration in our suspension as obtained from the ionic
conductivity. Given the additional presence of a charged substrate, it
might seem surprising that Eq.(\ref{Yukawa}) describes our data
successfully. However, it has been demonstrated experimentally
\cite{Grier} and theoretically \cite{Stillinger,Netz} that a
Yukawa-potential captures the leading order interaction also for
colloids close to a charged wall. A confining wall introduces only a
very weak (below 0.1 $k_{B}T$) correction due to additional dipole
repulsion. This correction is below our experimental resolution.
Repeating the two-body measurements with different laser intensities
(50mW to 600mW) yielded within our experimental resolution identical
pair potential parameters. This also demonstrates that possible
light-induced particle interactions (e.g. optical binding
\cite{Burns}) are neglegible. The approach of the third particle by
means of an additional optical trap could, in principle, lead to
additional light-induced interactions between the laser spot and the
two particles kept in the line trap. To exclude such effects, we
repeated the two-particle measurements and approached an empty trap
(without the third particle) to the line trap where the two particles
were fluctuating. Within our experimental resolution, we again
observed identical pair potentials, which suggests, that the
additional optical trap has no influence on the two particles in the
line trap.  When a third particle is present at a distance $d$ along
the perpendicular bisector of the scanned laser line (cf. inset of
Fig.\ref{photo}), the total interaction energy $U(r,d)$ is not simply
given by the sum of the pair-interaction potentials Eq.(\ref{Yukawa})
alone but also contains an additional term. Following the definition
of McMillan and Mayer \cite{McM}, $U(r,d)$ is given by
\begin{widetext}
\begin{eqnarray}
U(r,d) = u_{pair} (r_{12})+ u_{pair} (r_{13})+ u_{pair} (r_{23}) 
+ u_{123} (r_{12},r_{13},r_{23})\; , 
\label{u123}
\end{eqnarray}
\end{widetext}
with $u_{pair}(r_{ij})$ being the pair-potential between particles $i$
and $j$ as defined in Eq.(\ref{Yukawa}) and $u_{123}$ the three-body
interaction potential. Distances $r_{12}$, $r_{23}$ and $r_{13}$ are
the distances between the three particles which can, due to the chosen
symmetric configuration ($r_{23}\equiv r_{13}$), be expressed by the
two variables $r = r_{12}$ and $d=\sqrt{r_{13}^{2}-(r/2)^2}$. We have
followed the same procedure as described above for the case of two
particles. First, we have measured the probability distribution
$P(r;d)$ of the two particles in the laser trap with the third
particle fixed at distance $d$ from the trap.  Taking the logarithm of
$P(r;d)$ we extracted the total interaction energy $U(r,d)$
\cite{y-direction}.
The results are plotted as symbols in Fig.\ref{utot} for the distance
of the third particle $d = 4.1,$ 3.1, 2.5 and 1.6 $\mu$m,
respectively. As expected, $U(r,d)$ becomes larger as $d$ decreases
due to the additional repulsion between the two particles in the trap
and the third particle. In order to test whether the interaction
potential can be understood in terms of a pure superposition of
pair-interactions, we first calculated $U(r,d)$ according to
Eq.(\ref{u123}) with $u_{123} \equiv 0$. This was easily achieved
because the positions of all three particles were determined during
the experiment and the distance-dependent pair-potential is known from
the two-particle measurement described above (Fig.\ref{pairpot}). The
results are plotted as dashed lines in Fig.\ref{utot}. Considerable deviations
from the experimental data can be observed, in particular at smaller
$d$. These deviations can only be explained, if we take three-body
interactions into account. Obviously, at the largest distance, i.e. $d
= 4.1 \mu m$ our data are well described by a sum over pair-potentials
which is not surprising, since the third particle cannot influence the
interaction between the other two, if it is far away from both. In
agreement with theoretical predictions \cite{Carsten}, the three-body
interactions therefore decrease with increasing distance $d$.

\begin{figure}
 \includegraphics[width=0.75\textwidth]{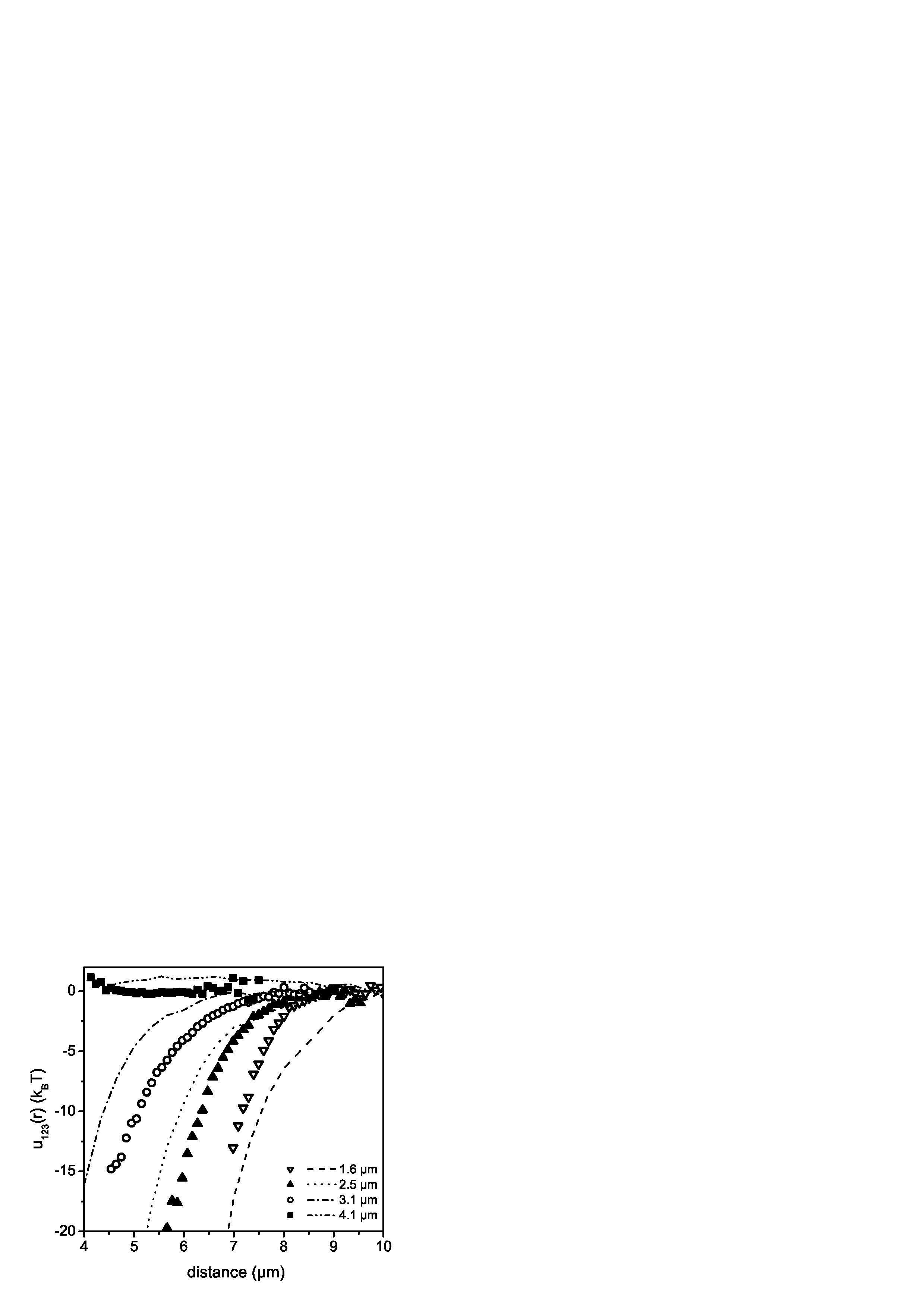}
 \caption{}
\label{u3}
\end{figure}

According to Eq.(\ref{u123}) the three-body interaction potential is
simply given by the difference between the measured $U(r,d)$ and the
sum of the pair-potentials (i.e. by the difference between the
measured data and their corresponding lines in Fig.\ref{utot}). The
results are plotted as symbols in Fig.\ref{u3}. It is clearly seen,
that in the case of charged colloids $u_{123}$ is entirely attractive
and becomes stronger as the third particle approaches. It is also
interesting to see that the range of $u_{123}$ is of the same order as
the pair-interaction potentials. It might seem surprising that it is
possible to sample the potential up to energies of 15$k_{B}T$, as
configurations of such a high energy statistically happen only with
very low probability. In this experiment we can choose the energetic
range of the potential we want to sample by adjusting the strength of
the line tweezers. The laser potential pushes the particles together,
which allows us to sample different ranges of the electrostatic
potential. Thus, to achieve a better resolution for smaller particle
separations (e.g. higher potential values), the strength of the line
tweezers had to be increased. The shape of the external potential
$u_{L}$ was independent of the strength of the laser beam (see
Fig.\ref{laserpot}) and the magnitude scaled linearly with the input
laser power. This allowed us to adjust the input laser intensity so as
to obtain a suitable particle separation range. The external potential
was obtained simply by scaling the Gaussian shown in Fig.\ref{laserpot}.

\section{Numerical calculations}

In order to get more information about three-body potentials in
colloidal systems, we additionally performed non-linear
Poisson-Boltzmann (PB) calculations, in a similar way as in
\cite{Carsten}. The PB theory provides a mean-field description in
which the micro-ions in the solvent are treated within a continuum
approach, neglecting correlation effects between the micro-ions. It
has repeatedly been demonstrated \cite{Groot,Levin} that in case of
monovalent micro-ions the PB theory provides a reliable description of
colloidal interactions. The interactions among colloids are, on this
level, mediated by the continuous distribution of the microions and
can be obtained once the local electrostatic potential due to the
microionic distribution is known. The normalized electrostatic
potential $\psi(x,y,z)$, which is the solution of the non-linear PB
equation,
\begin{eqnarray}
 \bnabla^{2} \psi(\vec{r}) & = & \kappa^{2} \sinh \psi(\vec{r})\;, 
\nonumber \\
{\bf n}\cdot\bnabla \psi & = & 4 \pi \lambda_{B} \sigma\;, \quad\quad
\vec{r} \;\rm{on}\, \rm{colloid}\, \rm{surface}\;,
\label{PBE}
\end{eqnarray}
describes the equilibrium distribution of the microions around a given
macroionic configuration. Here $\kappa$ is the inverse Debye screening
length , $\lambda_B$ the Bjerrum length ($\lambda_{B}=0.72$nm for
aqueous solutions at room temperature) and $\sigma$ is the surface
charge density on the colloid surface (constant charge boundaries are
assumed for all colloids in the system). ${\bf n}$ is the normal unit
vector on the colloid surface. We used the multi-centered technique,
described and tested in other studies \cite{EPL,JCP} to solve the PB
equation (\ref{PBE}) at fixed configurations of three colloids and
obtain the electrostatic potential $\psi(x,y,z)$, which is related to
the micro ionic charge density. Integrating the stress tensor,
depending on $\psi(x,y,z)$, over a surface enclosing one particle,
results in the force acting on this particle. First, we calculated how
the force $f_{12}$, and from it the pair-potential between two
particles, depend on the distance between isolated two particles.
Choosing the suitable bare charge on the colloid surface, we were able
to reproduce the measured pair-interaction in Fig.\ref{pairpot}. The
calculation of three-body potentials was then carried out by
calculating the total force acting on one particle in the line trap
(say, particle 1) in the presence of all three particles and
subtracting the corresponding pair-forces $f_{12}$ and $f_{13}$
obtained previously in the two-particle calculation. If there is any
difference between the force on particle 1 obtained from the full PB
solution for the three particle configuration and the sum of two
two-body forces, this difference is due to the three-body interactions
in the system. The difference is then integrated to obtain the
three-body potential. The results are plotted as dashed and dotted
lines in Fig.\ref{u3} and show qualitative agreement with the
experimental data. To account for the deviations from the experimental
data one has to take into account the following points: {\bf (i)}
there is a limited experimental accuracy to which the light potential
can be determined. The accuracy decreases with increasing laser
intensity (note that normalized potentials are plotted in
Fig.\ref{laserpot}). In the three-body experiments, due to the
presence of the third repulsive particle, a stronger light field is
needed and the experimental error in determining the light potential
is estimated to be around $\pm 1 k_{B}T$. Since we have to subtract
the light potential twice from the total potential to obtain the
three-body potential, this error doubles and we expect an error of
about $\pm 2 k_{B}T$ in the final result. {\bf (ii)} An error of about
$\pm 2k_{B}T$ should be expected in the numerically obtained
three-body potentials as well. {\bf (iii)} While in the numerical
calculation we assume identical colloidal spheres, in the experiment
small differences with respect to the size and the surface charge are
unavoidable. This effect, however, is rather small and leads to
deviations on the order of 5 percent of the total potential.  {\bf
  (iv)} The numerical calculations do not take into account any
effects which may be caused by the substrate. Although we expect such
effects to be rather small (similar to the contribution for the pair
potential) they can not completely ruled out.  Considering the above
mentioned uncertainties it should be emphasized that in particular the
sign and the order of magnitude of the calculated potential compares
well with our measured results. This strongly supports our
interpretation of the experimental results in terms of the three-body
interactions.

We have measured and calculated the three-body interaction on a
mesoscopic level, but since the colloidal interactions are mediated by
the microions distributed in an electrolyte around the colloids, it is
interesting to explore what happens on a microscopic level, i.e.,
what feature of the microscopic distributions lead to the observed
three-body interactions. Of course, it is not possible to observe the
microionic density experimentally, but in a Poisson-Boltzmann
simulation such an information is easily accessible. Since the
microion density depends monotonically on the electrostatic potential
$\psi(\vec{r})$, it is enough to compare the electrostatic potentials
to qualitatively discuss the microscopic picture. Of course, to a
large extent, the potential $\psi(\vec{r})$ around three particles is
just the superposition of potentials around individual particles, but
since the solutions of nonlinear equations are in principle not
superposable, we expect to find small differences. It is indeed these
small differences that are ultimately responsible for the three-body
interaction.

\begin{figure*}
 \includegraphics[width=1.1\textwidth]{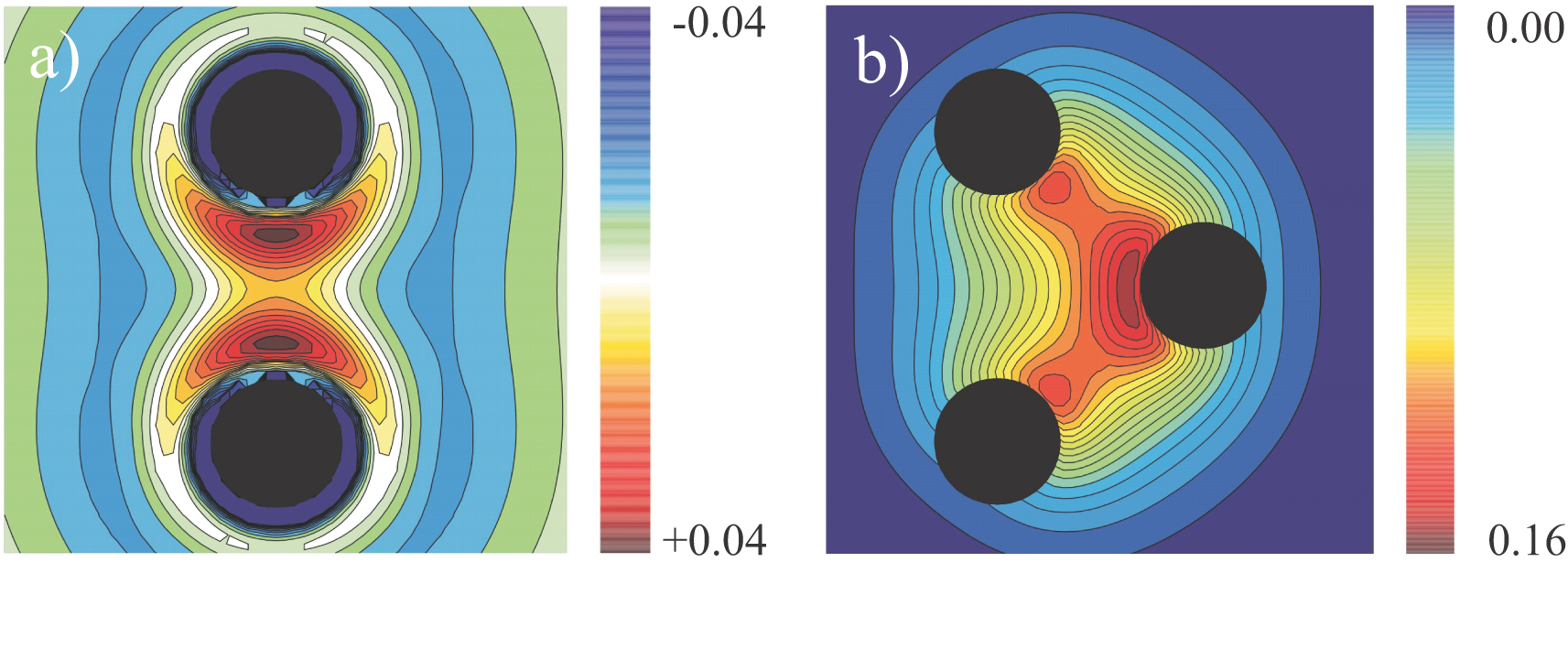}
 \caption{}
\label{contour}
\end{figure*}

We started by reconsidering the two-particle problem. First, we solved
the PB equation around a single isolated colloid to obtain the
one-body non-linear electrostatic potential $\psi^{1}(\vec{r})$. Next
we calculated the electrostatic potential $\psi^{2}(\vec{r})$ for two
colloidal particles at distance $r$ and compared this potential to the
superposition of two one-body potentials
$\psi^{1}_{1}(\vec{r})+\psi^{1}_{2}(\vec{r})$. The difference is shown
as a contour-plot in Fig.\ref{contour}a. It can be seen that
micro-ions are rearranged in a complex way between the colloids. There
is a weak additional polarization of the counter-ion cloud very close
to the particle surfaces not captured by superposing the one-body
potentials. However, all these effects are rather small and therefore,
except for very small particle separations r, the superposed solution
should still describe the two-body interactions with good accuracy.
Not so for three particles. We have compared a superposition of three
two-body electrostatic potentials with the correct non-linear
three-body electrostatic potential \cite{2b-3b}. The difference is
shown in Fig.\ref{contour}b. Obviously, differences are now much
larger than in Fig.\ref{contour}a. We notice that the counter-ion
cloud polarization close to the colloid surface is correctly taken
into account by two-body terms, while the ion distribution in the
region among the colloids is poorly described by adding up two-body
electrostatic potentials. There are fewer counter-ions in the region
among the colloids than a pair-wise description predicts. This
suggests that the entropy gained by removing some excessive
counterions (predicted by the superposition) from the inter-particle
space is larger than the positive energy difference due to less
efficient screening resulting from it. By integrating the potential
difference from Fig.\ref{contour}b, one recovers the attractive
three-body potential, already discussed, which is thus demonstrated to
be a consequence of the nonlinearity of the physical equations
governing the interactions in our system. The exact microscopic
explanation of the phenomenon, however, is still lacking and further
work is necessary to achieve it.

\section{Conclusions}

We have demonstrated that in case of three colloidal particles,
three-body interactions are attractive and of the same range as
pair-interactions. They present a considerable contribution to the
total interaction energy and must inevitably be taken into account.
Whenever dealing with systems comprised of many (i.e. more than three)
particles, in principle also higher-order terms have to be considered.
The relative weight of such higher-order terms depends on the particle
number density $\rho$ . While at low enough $\rho$ a pure pair-wise
description should be sufficient, with increasing density first
three-body interactions and then higher-order terms come into play. We
expect that there is an intermediate density regime, where the
macroscopic properties of systems can be successfully described by
taking into account only two- and three-body interactions \cite{Anti}.
Indeed liquid rare gases \cite{Jakse} and the island distribution of
adsorbates on crystalline surfaces \cite{Osterlund} are examples where
the thermodynamic properties are correctly captured by a description
limited to pair- and three-body interactions \cite{attractive-3b}. In
colloidal systems we have shown the three-body interactions to be
comparable in magnitude to the corresponding pair-interactions,
therefore we there expect large macroscopic three-body effects in this
intermediate density range. At even larger particle densities n-body
terms with $n>3$ have to be additionally considered, which may
partially compensate. Even in this regime, however, many-body effects
are not cancelled, but lead to notable effects, e.g. to a shift of the
melting line in colloidal suspensions, as recently demonstrated by PB
calculations \cite{EPL,JCP}.

With some effort it is in principle possible to proceed to measure the
higher order many-body terms and to study how the many-body expansion
converges. Work on four-body interactions is in progress.\\

\section{Acknowledgements}
Stimulating discussions with R. Klein, C. Russ and E. Trizac are
acknowledged. This work has been supported by the Deutsche
Forschungsgemeinschaft (Grants Be1788 and Gr1899).

\newpage

\section{Figure captions}

{\bf Fig.1}: Photograph of sample cell with two silica particles
confined to a light trap created by an optical tweezers and a third
particle trapped in a focused laser beam. The inset shows a schematic
drawing of the experimental geometry.\\

{\bf Fig.2}: The shape of the laser potential along the tweezers line
for three different laser intensities (symbols: triangles 100mW,
circles 200mW and squares 500mW); for better comparison all curves are
normalized to an intensity of 100mW. The Gaussian fit is plotted as a
solid line.\\

{\bf Fig.3}: Measured pair-interaction potentials $U(r)=u_{pair}(r)$
(symbols) in the absence of the third particle.  The data agree well
with a DLVO potential, Eq.(\ref{Yukawa}) (solid line). In the inset
the potential is multiplied by $r$ and plotted logarithmically, so
that the DLVO expression, Eq.(\ref{Yukawa}) transforms into a straight
line. From a fit we obtained the effective charge $Z^{*}\approx 6500$
and the screening length $\kappa^{-1} \approx
470$nm.\\

{\bf Fig.4}: Experimentally determined interaction energy $U(r)$
(symbols) for two particles in a line tweezers in the presence of a
fixed third particle with distance $d$ on the perpendicular bisector of
the line trap. For comparison the superposition of three
pair-potentials is plotted as lines. Symbols and lines are labelled by
the value of $d$.\\

{\bf Fig.5}: Three-body potentials for different $d$. Measured
three-body potentials indicated by symbols. The lines are three-body
potentials as obtained from the solutions of the nonlinear
Poisson-Boltzmann equation for three colloids arranged as in the
experiment. The parameters in the Poisson-Boltzmann calculation were
chosen so that the pair-interaction potentials were correctly
reproduced. Symbols and lines are labeled by the value of $d$.\\

{\bf Fig.6}: Contour plots of electrostatic potentials. {\bf a)}
Difference between the full electrostatic potential for two particles
and the superposition of two one-particle potentials. The distance
between the particles is $r=2.5 \mu$m. {\bf b)} Difference between the
full electrostatic potential for three particles and the superposition
of three two-particle potentials. The distance between particle 1 and
2 is $r=2.5 \mu$m and the position of the third particle is given by
$d=1.6 \mu$m, being the closest distance realized in the experiments.
The colour scales are in units of $k_{B}T$.\\

\newpage



\begin{thebibliography}{99}
\bibitem{Zsat} The saturated
effective charge of our particles is about 6900 for water at room
temperature.
\bibitem{y-direction} When
approaching the third particle, the two particles in the trap are
slightly displaced in the $y$ direction at small $r$. Accordingly, the
minimum energy configurations of the two particles are not on a
straight line as before. The most likely configuration at given
distance $r$ is $\left (x_{1}=r/2, y_{1}=y(r), x_{2}=-r/2,
y_{2}=y(r)\right )$ with $y(r)$ given by the measured particle
positions. Since we have the full knowledge of the two-dimensional
external laser potential, we can compute $u_{L}(r/2,y(r))$ for every
given configuration and use it in the Eq.(\ref{Pr}) instead of
$u_{L}(r/2,0)$.
\bibitem{2b-3b} In the configuration of three particles we have three
  pairs. We have first performed three Poisson-Boltzmann calculations
  with two particles (each pair isolated) to obtain the two-body
  electrostatic potentials $\psi^{2}_{ij}(\vec{r})$. We superposed
  those three solutions to obtain a prediction for the electrostatic
  potential around three particles
  $\tilde{\psi}(\vec{r})=\textstyle{\frac{1}{2}}\left
    (\psi^{2}_{12}(\vec{r})+
    \psi^{2}_{13}(\vec{r})+\psi^{2}_{23}(\vec{r})\right )$ without
  taking three-body effects into account. Then we performed the
  Poisson-Boltzmann calculation around three particles to obtain the
  correct electrostatic potential $\psi(\vec{r})$. The difference
  $\psi(\vec{r})-\tilde{\psi}(\vec{r})$ is plotted on
  Fig.\ref{contour}b and this is the microscopic origin of the
  measured three-body interaction $u_{123}(r,d)$.
\bibitem{attractive-3b} Three-body interactions are not always
  attractive. In case of van der Waals interactions, the AT-triple
  dipole interactions are predicted to be either positive or negative,
  depending on the configuration of the particles.
\end{thebibliography}
\end{document}